\documentclass[12pt]{iopart}

\usepackage[english]{babel} 	
\usepackage[T1]{fontenc} 			

\usepackage{graphicx}
\usepackage{subfig}

\usepackage[numbers,square,sort&compress,longnamesfirst]{natbib}

\usepackage{color}

\expandafter\let\csname equation*\endcsname\relax

\expandafter\let\csname endequation*\endcsname\relax

\usepackage{amsmath}

\usepackage[pdftex]{hyperref} 
\hypersetup{colorlinks,       
            linkcolor=black,  
            anchorcolor=black,
            citecolor=black,  
            filecolor=black,  
            menucolor=black,  
            urlcolor=black,   
	          bookmarksopen=true,    
	          bookmarksnumbered=true, 
	          pdftitle={Orbital motions and the Centre of Mass},
            pdfauthor={Andre Guerra, Paulo Carvalho},
            pdfsubject={Article on the orbital motions and the Centre of Mass},
            pdfstartview=FitH,
            pdfdisplaydoctitle=true,
						pdfpagemode=UseNone} 

\usepackage{booktabs}
\usepackage{multirow}

\newcommand{\vecu}[1]{\ensuremath{\widehat{#1}}}	
\newcommand{\td}{\ensuremath{\mathrm{d}}}        	
\newcommand{\unit}[1]{\ensuremath{\mathrm{#1}}}		

\begin{document}

\title[Orbital motions and the Centre of Mass]{Orbital motions of astronomical bodies and their Centre of Mass from different reference frames: a conceptual step between the Geocentric and Heliocentric models}

\author{Andr\'{e} G C Guerra$^1$\footnote{Corresponding author.} and Paulo Sime\~{a}o Carvalho$^{2,3}$}

\address{$^1$ Departamento de F\'{i}sica e Astronomia, Centro de F\'{i}sica do Porto, Faculdade de Ci\^{e}ncias, Universidade do Porto, Rua do Campo Alegre, s/n, 4169-007 Porto, Portugal}%

\address{$^2$ Departamento de F\'{i}sica e Astronomia, UEC, Faculdade de Ci\^{e}ncias, Universidade do Porto, Rua do Campo Alegre, s/n, 4169-007 Porto, Portugal}%

\address{$^2$ IFIMUP-IN, Rua do Campo Alegre, s/n, 4169-007 Porto, Portugal}%

\eads{\mailto{aguerra@fc.up.pt}, \mailto{psimeao@fc.up.pt}}

\vspace{10pt}

\begin{indented}
\item[]May 2016
\end{indented}

\begin{abstract}
The motion of astronomical bodies and the centre of mass of the system is not always well perceived by students.
One of the struggles is the conceptual change of reference frame, which is the same that held back the acceptance of the Heliocentric model over the Geocentric one.
To address the question, the notion of centre of mass, motion equations (and their numerical solution for a system of multiple bodies), and change of referential is introduced.
The discussion is done based on conceptual and real world examples, using the solar system.
Consequently, through the use of simple ``do it yourself'' methods and basic equations, students can debate complex motions, and have a wider and potentially effective understanding of physics.
\end{abstract}

\vspace{2pc}
\noindent{\it Keywords}: Reference Frame, Centre of Mass, Motion Equations, Numerical Solution, Geocentric Model, Heliocentric Model

\submitto{\PED}


\section{Introduction}
\label{sec:Introduction}

The notion of referential is one of the main causes why students have difficulties interpreting motions in astronomy.
This is not completely surprising, since the change of referential involves abstract reasoning, which is usually not completely developed in all population, especially young students~\cite{Shayer1983}.
Indeed, several examples of misconceptions were identified among students in many contents of physics and astronomy that require formal thinking~\cite{Driver2000,Arons1997}.

Looking back to the history of physics, the meticulous observations of nature by Aristotle (384 -- 322 BC) lead him to a philosophical description of the universe.
This was based on a simple, logic and very appealing common sense language.
For Aristotle, the Earth was the centre of the universe; the Earth was the only referential he knew.
Such geocentric description was lately improved by Ptolemy (c. 100 -- c. 170), in his work Almagest~\cite{Serzedello2012,Cotardiere2010,Ptolemy_Wiki2016,Almagest_Wiki2016}, where the motions of the planets, Moon and Sun around the Earth, were presented as a physical model supported by mathematical descriptions of the orbits.

The geocentric model had several advantages at that time: i. it did not contradict Aristotle's descriptions, ii. it was supported by the religious ideals of the middle age in Europe, and iii. it was based on a local frame of reference (the Earth) and followed a concrete reasoning.
That is why it prevailed for about 14 centuries.

When Copernicus (1473 -- 1543) took the (previously) rejected ideas of Aristarchus of Samos (c. 310 -- c. 230 BC) and proposed the Heliocentric model, he and his followers quickly realized they had two major issues to overcome: a huge and visible religious dispute, and a not-so-visible but still important cognitive barrier.
The cognitive barrier arose because it was necessary to put oneself outside the Earth to fully understand the observations taken from it.
It was this combination of factors that maintained the debate of Geocentric and Heliocentric-based models up to the moment when humankind sent satellites out of the Earth in the 20th century.

Nowadays, students still remain confused whenever changes of referential are needed to understand astronomical phenomena, such as moon phases, seasons, eclipses, or the motion of planets around a common centre of mass (CM).

Some simple but potentially effective ideas concerning the topic of referential were developed during the 11th Summer School of Physics of University of Porto for high secondary level students, and are presented here.
The concepts addressed are: CM of a system, trajectory of bodies in the CM referential, and absolute trajectory of the same bodies.
In this approach, some computing is used to simulate the trajectories of the bodies, and to discuss the implications of the initial conditions (mass and speed) on the description of the phenomena.

\section{Theory and Computational Method}
\label{sec:Theory}

Before diving into the subject of referential changes, it is important to introduce the theory needed for solving the problems numerically~\cite{Search1986}.

Consider a system composed by a number of distinct bodies, randomly distributed, orbiting each other.
For this setup, it is useful to define the location of the CM of the system.
The bodies' motion due to gravity forces can be computed with respect to a system centered on the CM, or to an inertial reference system.

\subsection{Centre of mass of a system}
\label{subsec:CM_System}

The CM of a multiple bodies system can be defined as the unique point, within the system, which can be used to describe the system's response to external forces.

The mass of the CM ($m_{CM}$) is the sum ($\sum$) of the mass of all $N$ bodies, each with mass $m_i$,
\begin{equation}
	m_{CM} \equiv \sum_{i=1}^{N} m_i.
	\label{eq:CM_mass}
\end{equation}

The position ($\vec{r}_{CM}$) and velocity ($\vec{v}_{CM}$) of the CM are related to each body location ($\vec{r}_{i}$) and velocity ($\vec{v}_{i}$), and can be computed by
\begin{align}
	\vec{r}_{CM} \equiv \frac{1}{m_{CM}} \sum_{i=1}^{N} m_i\vec{r}_i,
	\label{eq:CM_pos} \\
	\vec{v}_{CM} \equiv \frac{1}{m_{CM}} \sum_{i=1}^{N} m_i\vec{v}_i.
	\label{eq:CM_vel}
\end{align}

\subsection{Motion Equations}
\label{subsec:Motion_Equations}

To calculate the motion of the bodies in the system, Newton's second law has to be used,
\begin{equation}
	\vec{F}= m \vec{a},
	\label{eq:Newton}
\end{equation}
\noindent{}that relates the forces acting on a body ($\vec{F}$) and its acceleration ($\vec{a}$).

Admitting two bodies (of mass $m_1$ and $m_2$) in free space, the gravitational force acting on body 1 is
\begin{equation}
	\vec{F}_1 = G\frac{m_1 m_2}{r^2} \vecu{r},
	\label{eq:Newton_Gravity_r2}
\end{equation}
\noindent{}where $\vecu{r}$ is the unit vector that points from body 1 to body 2, and $G$ is the gravitational constant.
Using the notion of unit vector in equation~\ref{eq:Newton_Gravity_r2}, the result is
\begin{equation}
	\vec{F}_1 = G\frac{m_1 m_2}{r^3} \vec{r}.
	\label{eq:Newton_Gravity_r3}
\end{equation}

The acceleration of body 1 can thus be obtain from both equations~\ref{eq:Newton} and~\ref{eq:Newton_Gravity_r3},
\begin{equation}
	\vec{a}_1 = G\frac{m_2}{r^3} \vec{r}.
	\label{eq:Newton_motion}
\end{equation}
\noindent{}The same can be applied to body 2, just changing $m_2$ to $m_1$ and inverting the direction of $\vec{r}$ in equation~\ref{eq:Newton_motion}.

\subsubsection{Numerical Solution}

There are many methods to solve equation~\ref{eq:Newton_motion}.
Here a numerical one is used because a general solution for a system of multiple bodies is searched, and there is no analytical solution.

The simplest numerical method is the Euler one (and can even be applied with a spreadsheet editor~\cite{Slegr2012}).
It states that for a general function $f(t)$, it is possible to compute its value at an instant $t^{\,\{n+1\}}$, with an initial value $f^{\,\{n\}}$ (at instant $t^{\,\{n\}}$), using the relation
\begin{equation}
	f^{\,\{n+1\}} = f^{\,\{n\}} + \Delta t \left(\frac{\td f}{\td t}\right)^{\{n\}},
	\label{eq:Euler}
\end{equation}
\noindent{}where $\left(\td f/\td t\right)^{\{n\}}$ is the function derivative at instant $t^{\,\{n\}}$, and $\Delta t = t^{\,\{n+1\}} - t^{\,\{n\}}$.

Applying this method to the multiple body system, and knowing that $\td \vec{v}/\td t = \vec{a}$, the velocity of a body can be estimated (according to equation~\ref{eq:Euler}) by
\begin{equation}
	\vec{v}^{\,\{n+1\}}_i = \vec{v}^{\,\{n\}}_i + \Delta t \vec{a}^{\,\{n\}}_i,
	\label{eq:Euler_vel}
\end{equation}
\noindent{}where from equation~\ref{eq:Newton_motion}, the acceleration for body $i$, subject to the gravity of the others, is
\begin{equation}
	\vec{a}^{\,\{n\}}_i = \sum_{j=1,j \neq i}^{N}\frac{G m_j}{||\vec{r}^{\,\{n\}}_{ij}||^3}\vec{r}^{\,\{n\}}_{ij},
	\label{eq:Euler_acc}
\end{equation}
\noindent{}where $\vec{r}_{ij}$ is the vector pointing from body $i$ to body $j$. Applying the same scheme to the position ($\td \vec{r}/\td t = \vec{v}$), it yields
\begin{equation}
	\vec{r}^{\,\{n+1\}}_i = \vec{r}^{\,\{n\}}_i + \Delta t \vec{v}^{\,\{n\}}_i.
	\label{eq:Euler_pos}
\end{equation}

The Euler method is considered a first-order numerical method, where the global error is proportional to the step size ($\Delta t$).
As the examples latter discussed involve astronomical bodies (where the steps used can be of the order of days), an adjustment to the above formulae can be made, to have a greater accuracy.
One simple adjustment is to compute the position with a second order derivative, using the following equation,
\begin{equation}
	\vec{r}^{\,\{n+1\}}_i = \vec{r}^{\,\{n\}}_i + \vec{v}^{\,\{n\}}_i\Delta t + \frac{1}{2}\vec{a}^{\,\{n\}}_i \left(\Delta t\right)^2.
	\label{eq:Euler_pos_2nd}
\end{equation}

The above sequence of equations (\ref{eq:Euler_vel},~\ref{eq:Euler_acc} and \ref{eq:Euler_pos_2nd}) is repeated successively, until a pre-determined final instant is reached.
A way to verify how much error is introduced by the numerical solution, is to calculate the variation in the total mechanical energy~\cite{Search1989}.

\subsection{Frames of Reference}
\label{subsec:Frames_Reference}

All previous equations describe the motion in an inertial reference system.
This frame does not have translational or rotational acceleration, relative to the ``fixed stars''~\cite{Curtis2010}.
The Heliocentric frame is an example of an inertial referential for the Earth-Moon system, where the Sun is at the origin of the reference frame.

Sometimes, the motion of the bodies can be simplified when described at the CM reference system (where the origin is at the CM).
In this case, it is necessary to impose mathematically that $\vec{v}_{CM}$ is zero, as well as $\vec{r}_{CM}$.
This implies that the velocities and position of the bodies in the CM referential, respectively $\vec{v}_i^{\,'}$ and $\vec{r}_i^{\,'}$, are related to the absolute ones by,
\begin{align}
	\vec{v}_i^{\,'} = \vec{v}_i - \vec{v}_{CM},
	\label{eq:vel_CMframe} \\
	\vec{r}_i^{\,'} = \vec{r}_i - \vec{r}_{CM}.
	\label{eq:pos_CMframe}
\end{align}

This change of reference frame is a simple translation of the inertial reference frame, and can be applied to any general point by replacing $\vec{v}_{CM}$ and $\vec{r}_{CM}$ by the respective velocity and position of that point.
No rotation is introduced here.

\section{Examples}
\label{sec:Examples}

With the equations presented in the previous section, several multiple bodies systems can be solved.
Some examples, explored with students at the 11th Summer School of Physics, are presented here.

The first one is a generic two body system (which can be two stars or two planets).
The next ones correspond to real examples in the solar system, which include the Sun, and the planets Earth, Moon, Mars and Jupiter.
To deal with real examples, the Jet Propulsion Laboratory HORIZON Ephemerides, corresponding to 11th April 2016, were used to extract the bodies' real position and velocity~\cite{JetPropulsionLaboratory2016}.

\subsection{Example 1 -- Two Bodies General System}
\label{subsec:Two_Bodies}

The CM of a two bodies system is always lying between them, closer to the more massive body, or in the middle if they have the same mass.
Admitting one of the bodies has an initial velocity (table~\ref{tab:Initial_Cond}), students easily understand that both will start moving around each other, due to the gravitational interaction between them.
The trajectory of the bodies can be computed with a simple spreadsheet editor, as can be consulted in reference~\cite{Guerra2016_Excel}.
\begin{table}[htbp]
  \centering
  \caption{Two bodies system initial conditions}
	\label{tab:Initial_Cond}
    \begin{tabular}{cccc}
    \toprule
			Body  & Mass [$\times 10^{22}\,\unit{kg}$]	& Position ($x$, $y$) [$\times 10^{3}\,\unit{km}$]	& Velocity ($v_x$, $v_y$) [\unit{km/s}] \\
    \midrule
			1     & 59.72 															& (0, 0) 																						& (0, 0) \\
			2     & 7.346 															& (357.0, 0) 																				& (0, 0.2500) \\
    \bottomrule
    \end{tabular}
\end{table}

\begin{figure}[!htb]
  \centering
  \includegraphics[width=0.6\textwidth]{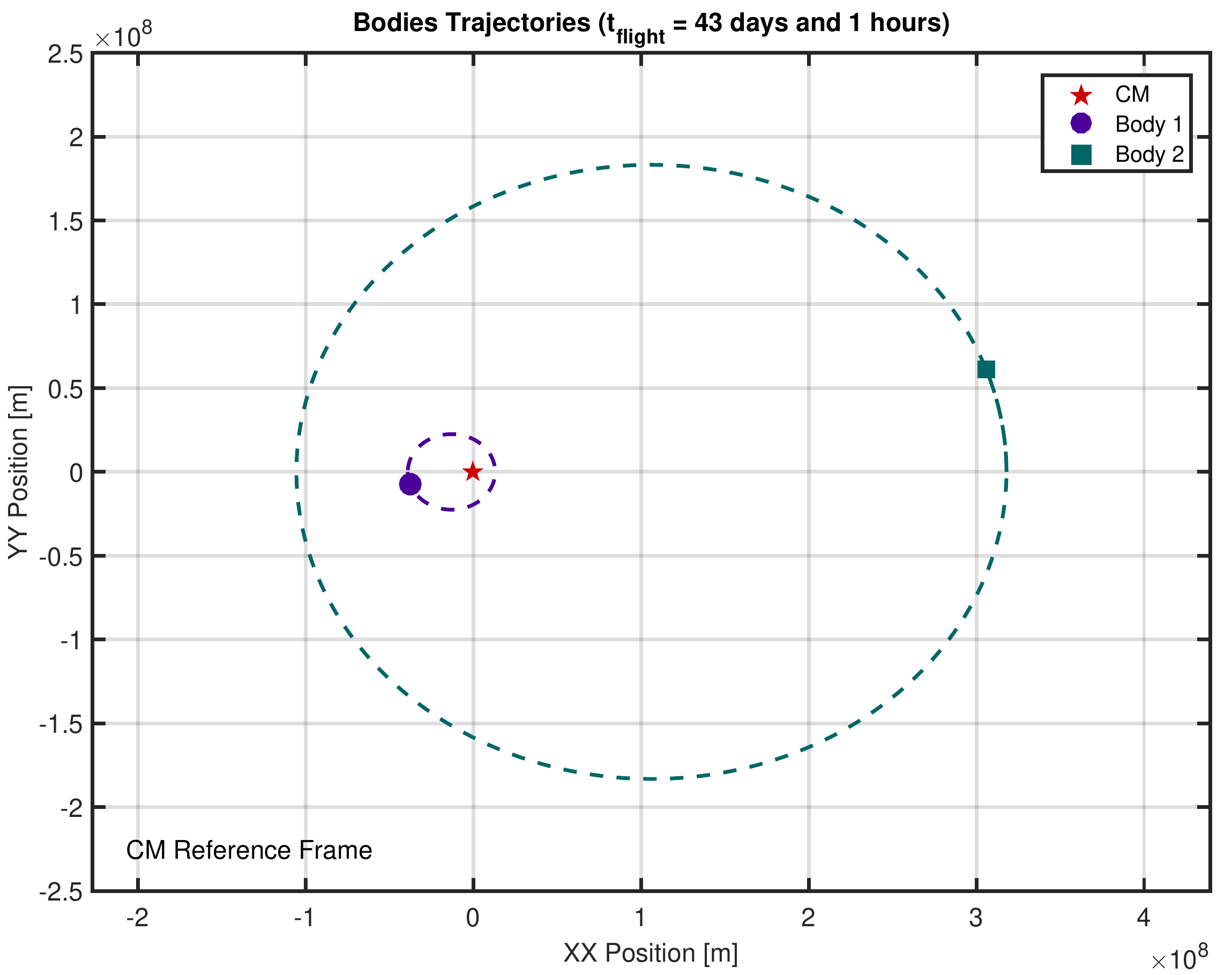}
  \caption{Two body system motions on the CM reference frame. Data was computed for a time interval of 90 days (only 43 days are shown here).}
  \label{fig:2Bodies_CMR}
\end{figure}

\begin{figure}[!htb]
  \centering
  \includegraphics[width=0.6\textwidth]{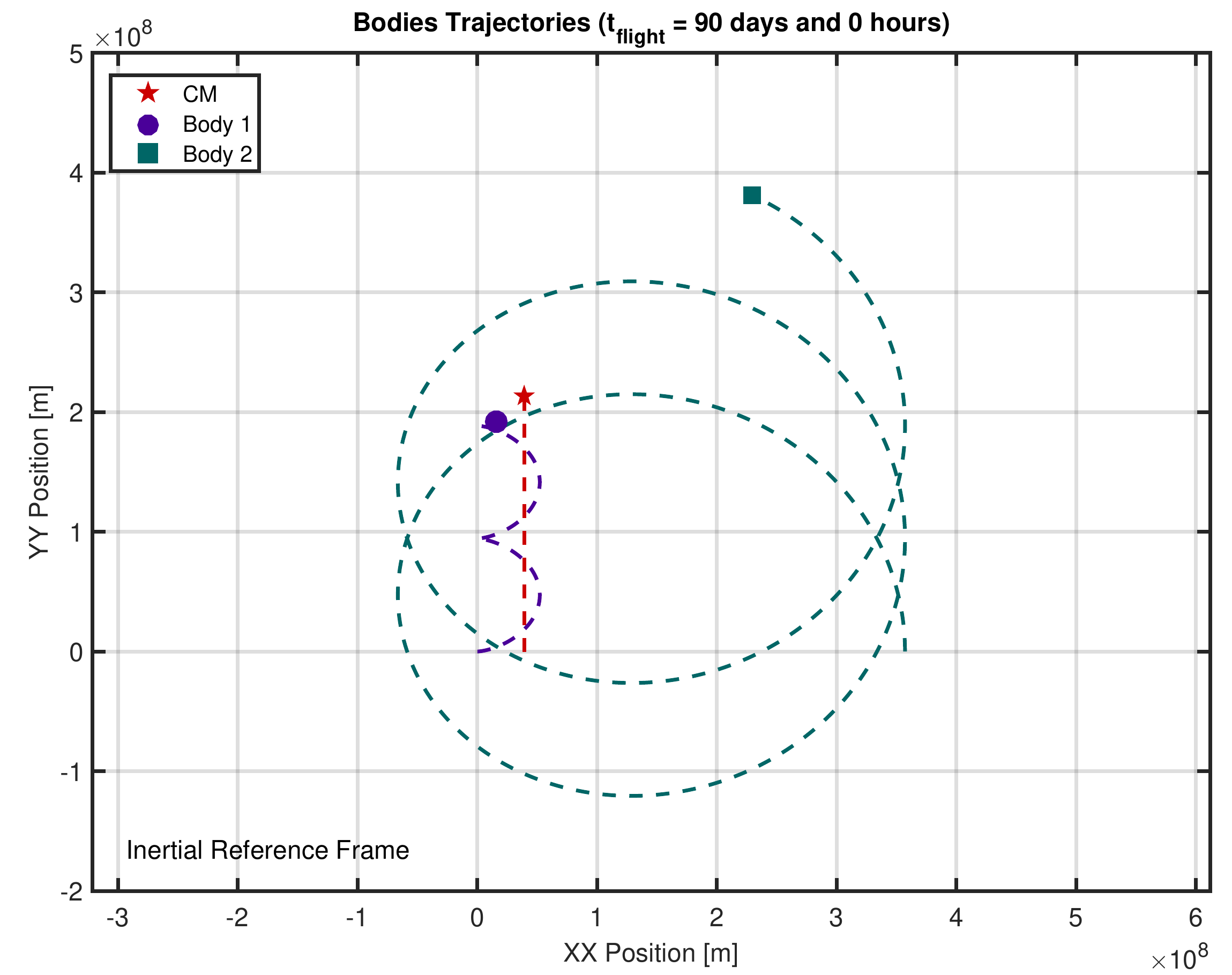}
  \caption{Two body system motions on an inertial reference frame. Data was computed for a time interval of 90 days.}
  \label{fig:2Bodies_Final_IR}
\end{figure}

It is therefore not surprising that students conclude the both bodies describe, in the CM reference system, closed elliptical trajectories around a common point, the CM (figure~\ref{fig:2Bodies_CMR}).
However, as the bodies' motion influence the velocity of the CM, the result in the inertial referential is a helical-like motion of the lower massive body, while the more massive one experiences a composition of translation in space and rotation around the CM (figure~\ref{fig:2Bodies_Final_IR}).

Such weird trajectories are counter-intuitive and provide a good first discussion about why we observe unlike trajectories in different reference systems.

\pagebreak 

\subsection{Example 2 -- Earth-Moon System}
\label{subsec:Earth_Moon}

The Earth-Moon system has some similarities to the previous example, but now the Earth is about ten times more massive (the Moon's mass is 1.23\% of the Earth).
Therefore, the CM of this system is much closer to the Earth.

\begin{figure}[!htb]
	\centering
	\subfloat[Earth and Moon seen from the CM]{
	\includegraphics[width=0.48\textwidth]{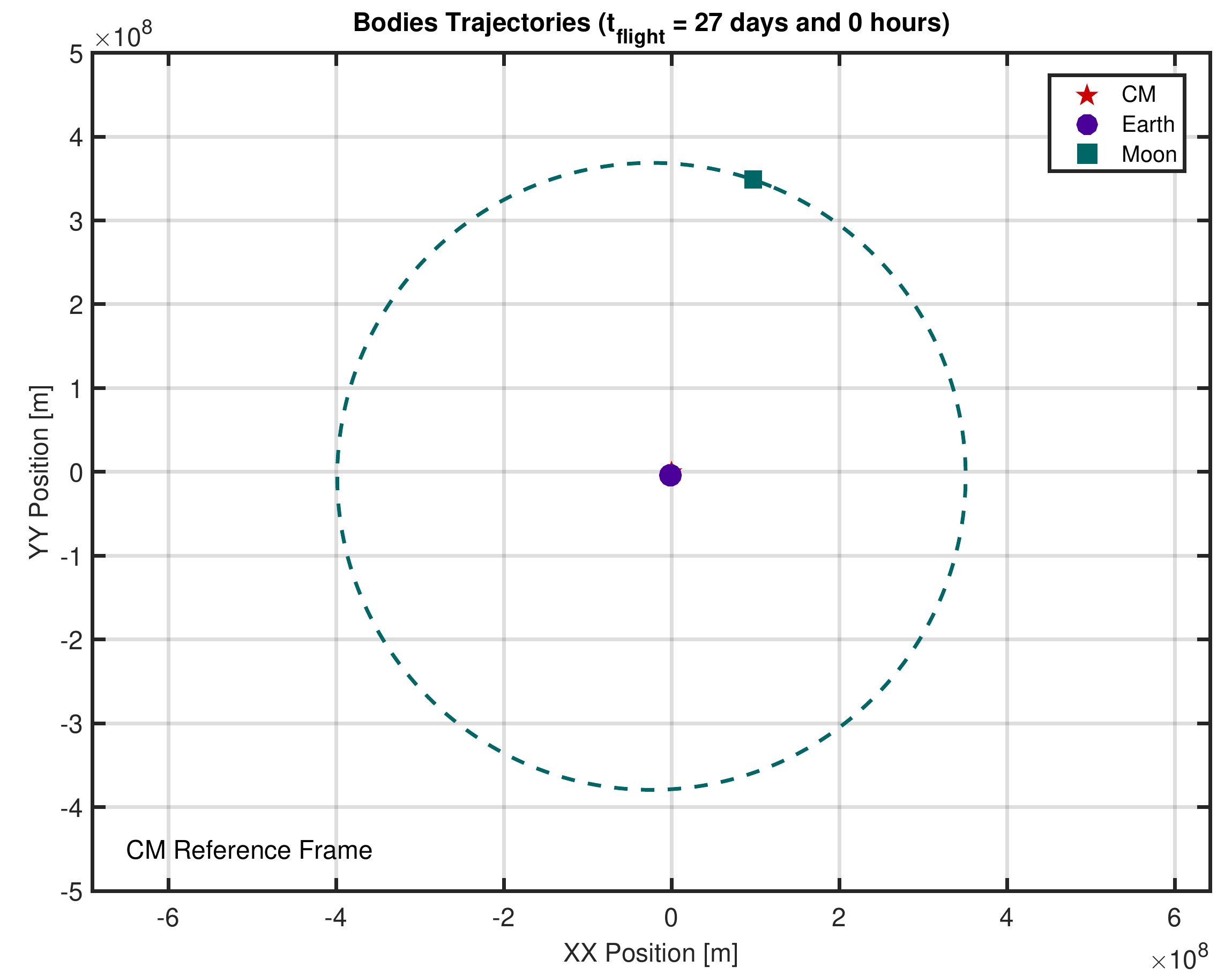}
	\label{fig:Earth_Moon_Global}}
	\subfloat[The CM seen from Earth]{
	\includegraphics[width=0.48\textwidth]{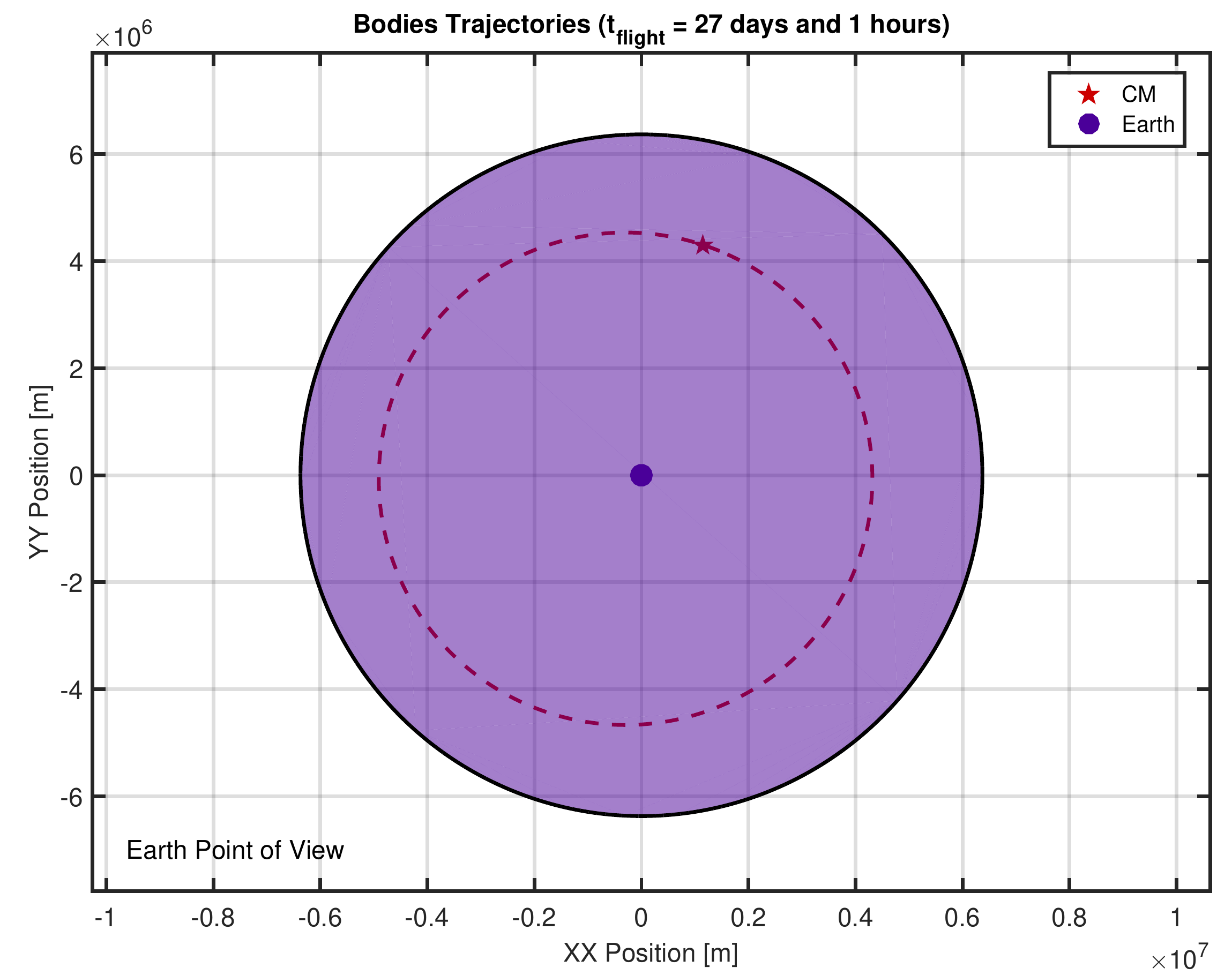}
	\label{fig:Earth_Moon_Ezoom}}
	\caption{The Earth-Moon system motion. In (a) the Earth graphically overlaps the CM. Data was computed for a time interval of 90 days (only 27 days are shown here).}
	\label{fig:Earth_Moon}
\end{figure}

In the CM referential, the Moon describes an elliptical motion, whereas the Earth does not seem to move (figure~\ref{fig:Earth_Moon_Global}).
However, a closer observation of the Earth reveals that in fact it moves, but the CM of the system is always located inside the planet (figure~\ref{fig:Earth_Moon_Ezoom}).

Because the CM lies inside the Earth, the first astronomers had the perception that the Moon orbited the Earth and not a common point (the CM)!
That only happened because their astronomical observations were taken from the Earth local referential!

\subsection{Example 3 -- Sun-Earth-Moon System}
\label{subsec:Sun_Earth_Moon}

Moving from the Earth-Moon system to the more general Sun-Earth-Moon system, we have to considerer that the Earth, in fact, orbits around the Sun and therefore has already a velocity in space.

At this stage, students are already aware that the most massive body (the Sun) is affected by the gravitational forces of the other two (Earth and Moon).
Thus, taking into account the previous example, they can predict that the centre of the Sun experiments some motion on the inertial reference system, which means that the solar system as a whole moves in space!
What they really may find surprising is that this motion is very small when compared to the Sun dimension (which has about 99.8\% of the total solar system mass), and this is why we state that the planets orbit only around the Sun in the solar system.

\begin{figure}[!htb]
  \centering
  \includegraphics[width=0.6\textwidth]{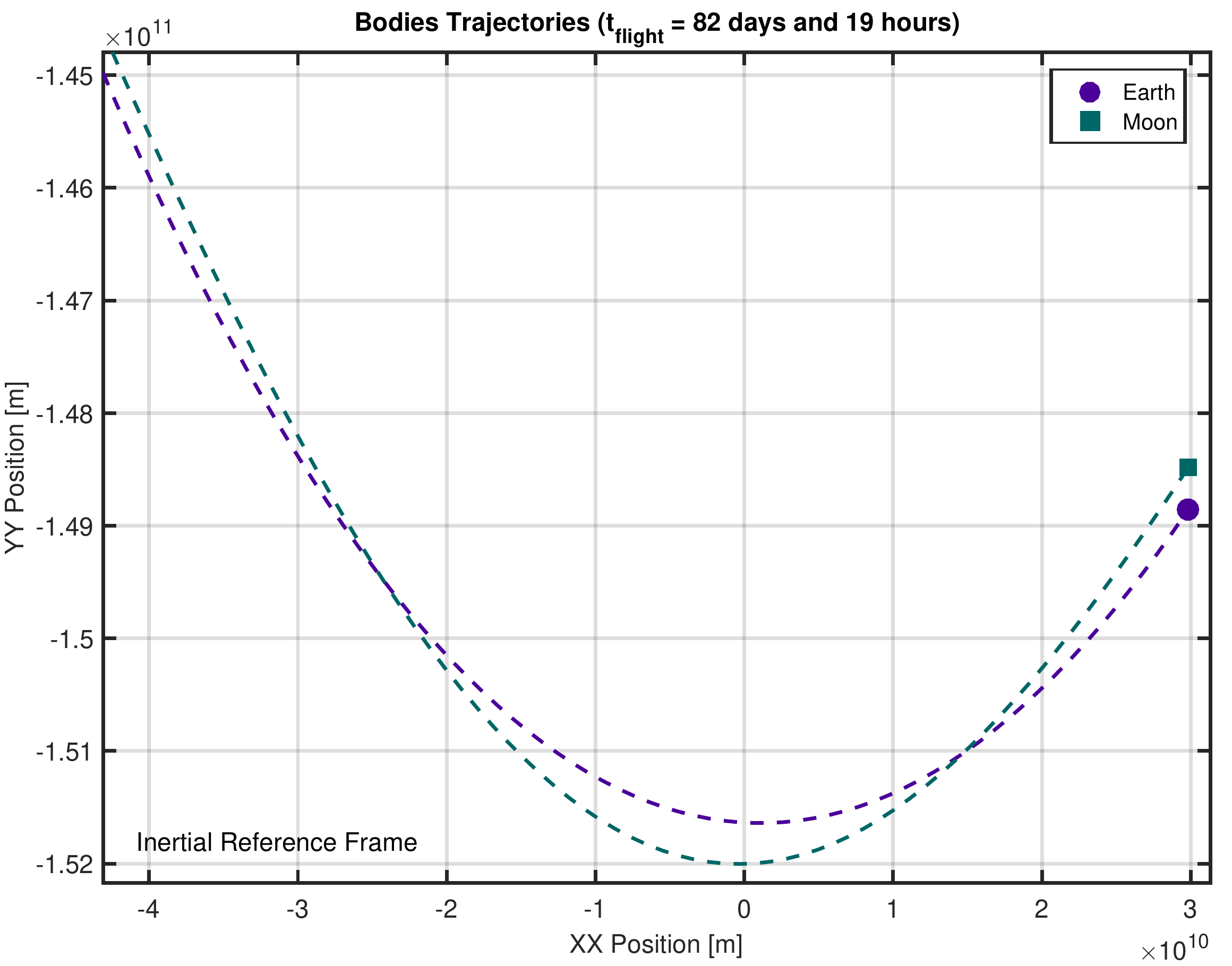}
  \caption{A close-up of the Earth-Moon system, orbiting the Sun, seen from an inertial reference frame. Data was computed for a time interval of 400 days (only a few days are shown here).}
  \label{fig:EMS_IR_EMzoom}
\end{figure}

The relative motions of the Earth and Moon can also be observed from the CM of this system of bodies.
Figure~\ref{fig:EMS_IR_EMzoom} shows a zoom of the trajectories of these bodies while they orbit the Sun.
Students can see that the Moon's trajectory is a composition of its orbit around the Earth, and the Earth around the Sun.
Therefore, the Moon intersects the trajectory of the Earth each time it completes an orbital revolution, resulting in an oscillating motion around the Earth trajectory.

The final stage of this topic is to let students understand why the (apparent) retrograde motion is seen from the Earth.
For this purpose, a more complex system must be projected.

\subsection{Example 4 -- More Complex Systems}
\label{subsec:More_Complex_Systems}

The principles for the mechanics of a more complex system do not differ from those in the previous examples.

\begin{figure}[!htb]
  \centering
  \includegraphics[width=0.6\textwidth]{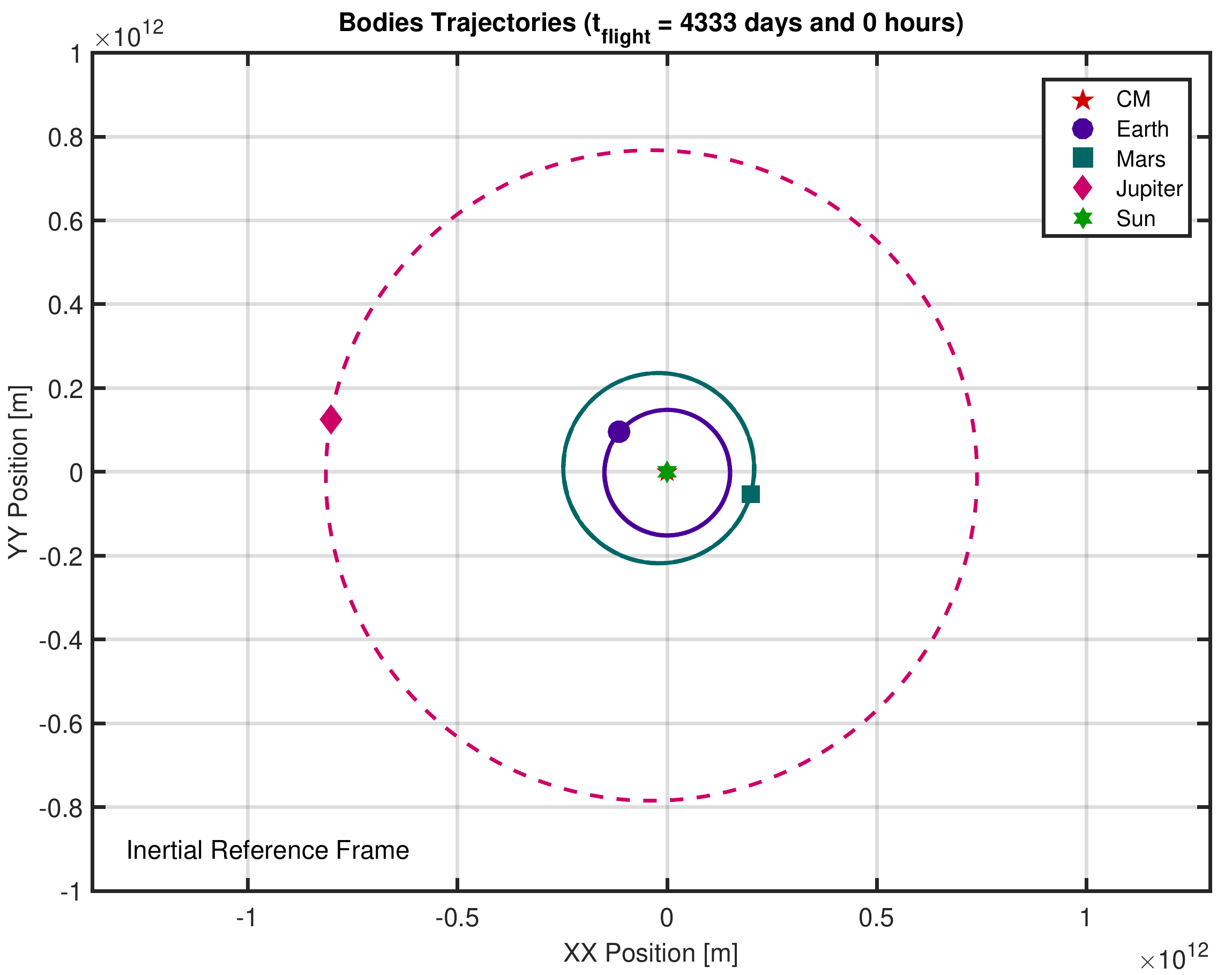}
  \caption{The motion of the Sun-Earth-Mars-Jupiter system, as seen from an inertial reference frame. In this image the Sun graphically overlaps the CM. Data was computed for a time interval of 4333 days.}
  \label{fig:EMaJS_Final_IR}
\end{figure}

The example considered here is the motions of the Earth, Mars and Jupiter around the Sun.
The resulting trajectories of these bodies on an inertial reference frame can be seen in figure~\ref{fig:EMaJS_Final_IR}.

Even though they influence each other, through gravitational forces, the planets follow almost circular orbits around the Sun, as was expected.
However, this is only true when seen from an inertial reference frame, or the Sun since his motion is negligible when compared to the planets' ones.

To evaluate the (apparent) motion of those planets as seen from a local referential (the Earth), students only need to apply equations (\ref{eq:vel_CMframe}) and (\ref{eq:pos_CMframe}), replacing $\vec{v}_{CM}$ and $\vec{r}_{CM}$ by $\vec{v}_{Earth}$ and $\vec{r}_{Earth}$, respectively, to obtain the velocity and position in the Earth reference frame.

\begin{figure}[!htb]
  \centering
  \includegraphics[width=0.6\textwidth]{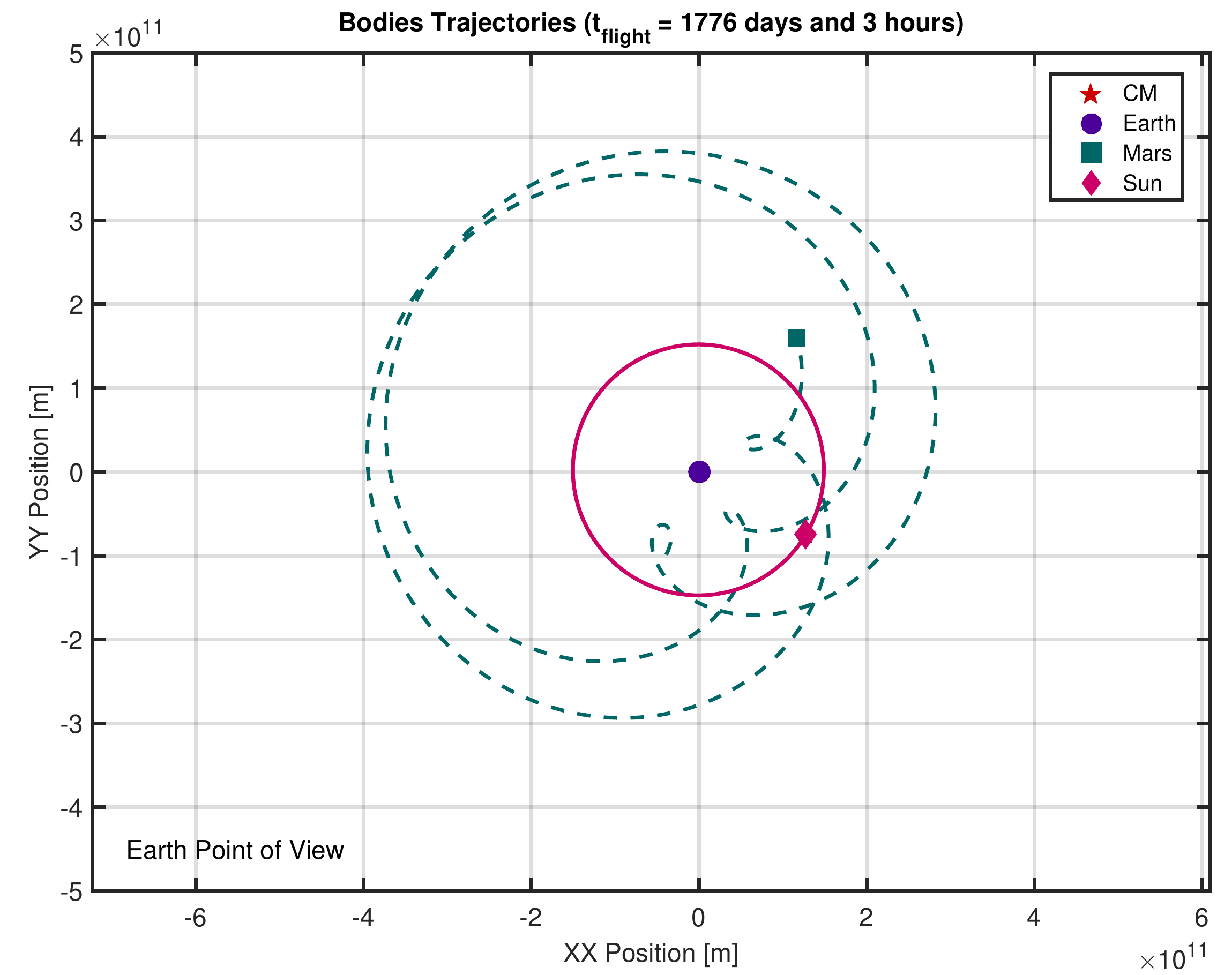}
  \caption{The motion of the Sun-Earth-Mars system, as seen from the Earth. In this image the Sun graphically overlaps the CM. Data was computed for a time interval of 2748 days (only 1776 days are shown here).}
  \label{fig:EMaS_ER}
\end{figure}

\begin{figure}[!htb]
  \centering
  \includegraphics[width=0.6\textwidth]{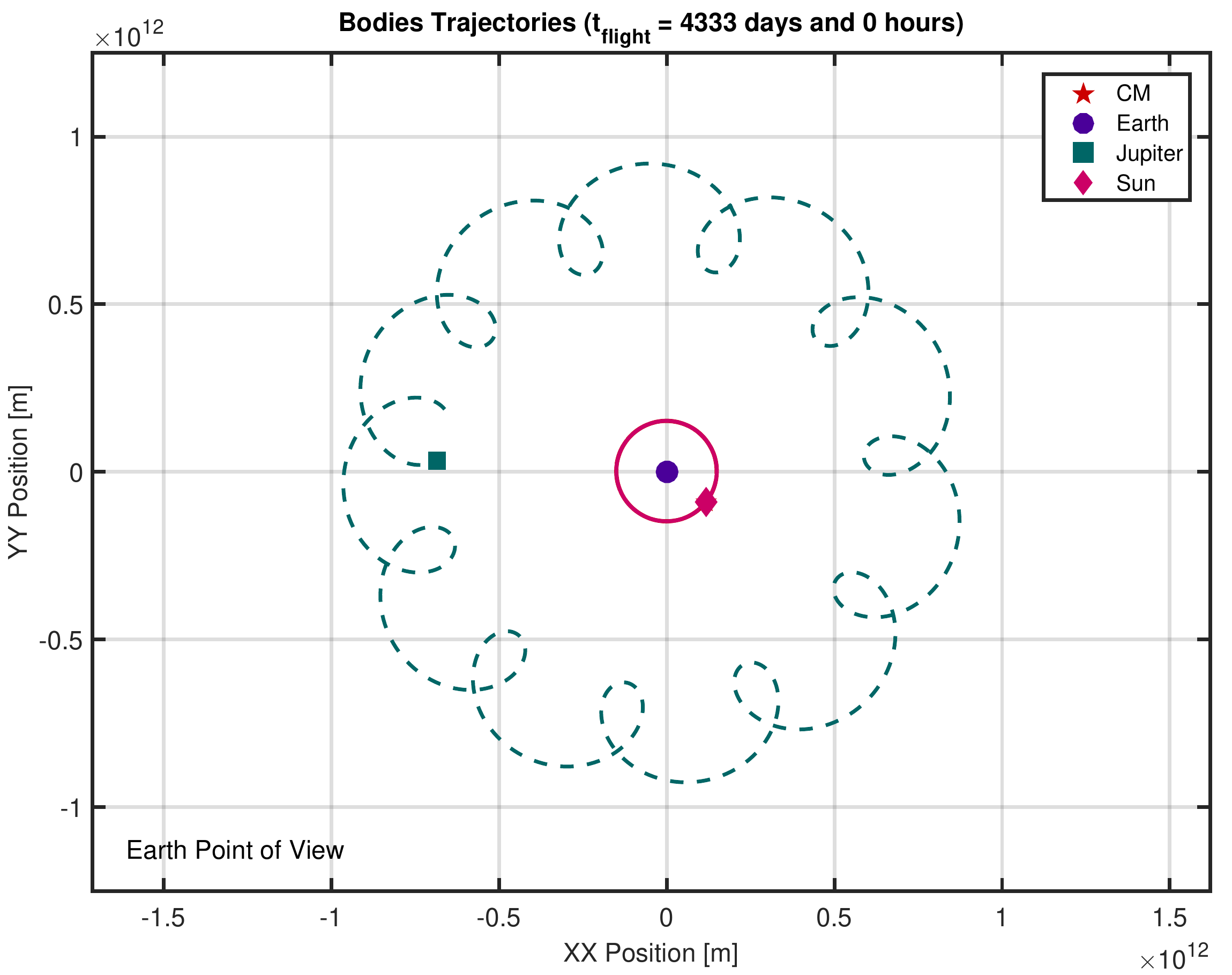}
  \caption{The motion of the Sun-Earth-Jupiter system, as seen from the Earth. In this image the Sun graphically overlaps the CM. Data was computed for a time interval of 4333 days.}
  \label{fig:EJS_Final_ER}
\end{figure}

Plots for the case of Sun-Earth-Mars and Sun-Earth-Jupiter systems are given in figures~\ref{fig:EMaS_ER} and~\ref{fig:EJS_Final_ER}, respectively.

The loops observed on Mars and Jupiter trajectories correspond to apparent retrograde motions. In ancient times these loops lead Ptolemy to formulate the epicycle theory for the Geocentric model, as an attempt to explain the phenomena observed for the five known planets at that time: Mercury, Venus, Mars, Jupiter and Saturn.
Instead of making the Geocentric model more accepted, the complexity of such theory weakened it.

Students can thus realise that the Heliocentric model proposed by Copernicus offers a simpler explanation of the planets' motion, clearly revealing why the retrograde motions are observed from the Earth.

To obtain all these plots an educational program~\cite{Guerra2016_OrbMotions} was developed (for a x64 Windows operating system), based on MATLAB software.
This program allows to see step by step the bodies' motions, which makes it easier and more engaging for students to understand, especially when they observe Mars and Jupiter making retrograde motions, as seen from the Earth.

\section{Conclusion}
\label{sec:Conclusion}

Students usually give very little importance to the meaning of Centre of Mass of a system, because they tend to confine it to the particle model problems.
The approach described in the context of astronomy gives a new and wider perspective of understanding of the role of CM in the dynamics of a system of bodies.
In particular, it promotes students' reasoning about how the trajectories of bodies can be seen in different reference frames, which is essential to understand natural phenomena such as the Coriolis effect or the Earth tides.

This approach also allows students to realise that apparently complex motions can be explained with basic and simple laws.
In fact, the most interesting about this approach is that students, with very simple mathematics but significant physics reasoning, can obtain these results by themselves, only considering gravitational interactions and no pre-established formulas.

The use of a simple spreadsheet where a convenient, but accessible, manipulation of equations can be made to solve the problems numerically, is a very engaging strategy that involves actively students in team-based learning and results in an effective understanding of physics.

\section*{Acknowledgements}
 The work of A. G., and P. C., is supported by the Funda{\c c}\~ao para a Ci\^encia e a Tecnologia (Portuguese Agency for Research) fellowship
 PD/BD/113536/2015, and Project UID/NAN/50024/2013, respectively.

\bibliographystyle{iopart-num}
\bibliography{ArticleCM_Bib}

\end{document}